\documentclass[12pt]{iopart}
\usepackage{iopams}

\newcommand{\bea}{\begin{eqnarray}}
\newcommand{\eea}{\end{eqnarray}}
\newcommand{\beq}{\begin{equation}}
\newcommand{\eeq}{\end{equation}}

\newcommand{\eqref}[1]{(\ref{#1})}

\def\edth{{\rlap{$\partial$}\raise0.3em\hbox{$-$}}}

\begin{document}

\title[Gravitational wave extraction from binary simulations]
{A note on gravitational wave extraction from binary simulations}

\author{
Hiroyuki Nakano$^{1,2}$
%Carlos O. Lousto$^{2}$,
%James Healy$^{2}$,
%\\
%Yosef Zlochower$^{2}$
}

\address{$^1$Department of Physics, Kyoto University,
Kyoto 606-8502, Japan.}

\address{$^2$Center for Computational Relativity and Gravitation,
and School of Mathematical Sciences, Rochester Institute of
Technology, Rochester, New York 14623, USA.}

%%%%%%%%%%%%%%%%%%%%%%%%%%%%%%%%%%%%%%%%
\begin{abstract}

In previous works, we developed 
a perturbative extrapolation formula
to obtain gravitational waves at infinity
from binary black hole simulations,
based on the Regge-Wheeler-Zerilli formalism.
In practice, this formula was basically derived
in a background flat spacetime.
We derive here an estimation of errors and
an improved extrapolation formula 
by using the Teukolsky perturbative equation
for a background Kerr spacetime.
The improved formula requires knowledge
about a mass and spin of the background spacetime.
\end{abstract}


\maketitle
%%%%%%%%%%%%%%%%%%%%%%%%%%%%%%%%%%%%%%%%

%%%%%%%%%%%%%%%%%%%%%%%%%%%%%%%%%%%%%%%%
\section{Introduction}\label{sec:intro}
%%%%%%%%%%%%%%%%%%%%%%%%%%%%%%%%%%%%%%%%

In numerical relativity (NR)
gravitational waveforms from binary black hole (BH) systems~\cite{Pretorius:2005gq, Campanelli:2005dd, Baker:2005vv}
are one of the most important output from the simulation.
In~\cite{Lousto:2010qx}, 
we have proposed a perturbative extrapolation formula for the Weyl scalar $\psi_4$: 
\bea
\left. r\,\psi_4^{\ell m}\right|_{r=\infty} = r\,\psi_4^{\ell m}(t,r)
- \frac{(\ell -1)(\ell +2)}{2\,r} \int dt \,[r\,\psi_4^{\ell m}(t,r)]
+ O(r^{-2}) \,,
\eea
decomposed
in the spin-weighted spherical harmonics (${}_{-2}Y_{\ell m}$)
expansion to obtain gravitational waves at infinity,
from a finite extraction radius $r$.
Here, $r$ denotes an approximate areal radius,
and $\psi_4^{\ell m}(t,r)$ is the $(\ell,\,m)$ mode of $\psi_4$
at finite radius $r$.
We see that the above equation is obtained from the wave equation
in a flat spacetime.
It should be noted that gravitational waveforms ${h}_{+/\times}$
are related to $\psi_4$ as $\psi_4 = \ddot{h}_{+} - i\,\ddot{h}_{\times}$
and that this is true only as $r \to \infty$. 

The above formula works well and has been confirmed 
for example in~\cite{Babiuc:2011qi}
by comparison with a characteristic evolution code 
to obtain the gravitational waveform at null infinity.
Also, the extrapolation formula has been used in~\cite{Kyutoku:2014yba}
for NR simulations of neutron star binaries.

Here, we revisit the extrapolation formula in~\cite{Lousto:2010qx} and improve it.
This paper is organized as follows.
Section~\ref{sec:pert} gives a brief summary of the derivation of
the extrapolation formula
and the effect on waveforms.
In Section~\ref{sec:Lazarus}, the original formula developed
in the BH perturbation approach~\cite{Regge:1957td,Zerilli:1971wd,Teukolsky:1973ha}
is improved by use of the Weyl scalar $\psi_4$ in NR.
The improved extrapolation formula requires information of 
a mass and spin of the background Kerr spacetime.
We discuss a proper choice of the mass and spin for the improved formula,
and the difference between
the original and improved formulas in section~\ref{sec:diss}.
All throughout, we use geometric units, with $G=c=1$.

%%%%%%%%%%%%%%%%%%%%%%%%%%%%%%%%%%%%%%%%
\section{Perturbative extrapolation}\label{sec:pert}
%%%%%%%%%%%%%%%%%%%%%%%%%%%%%%%%%%%%%%%%

In the Regge-Wheeler-Zerilli (RWZ) formalism~\cite{Regge:1957td,Zerilli:1971wd},
we consider gravitational waves, 
\bea
\fl
h_{+} - i\,h_{\times} =
\sum_{\ell \geq 2,\,|m| \leq \ell} \frac{\sqrt{(\ell-1)\ell(\ell+1)(\ell+2)}}{2r} 
\left(
\Psi_{\ell m}^{\rm (even)}(t,r) - i\, \Psi_{\ell m}^{\rm (odd)}(t,r)
\right)
{}_{-2}Y_{\ell m}
\,,
\label{eq:RWZwaves}
\eea
where $\Psi_{\ell m}^{\rm (even)}$ and $\Psi_{\ell m}^{\rm (odd)}$ denote
the even and odd parity wave functions, respectively.
The wave functions satisfy the RWZ equations,
\bea
\left[-\frac{\partial^2}{{\partial t}^2} 
+ \frac{\partial^2}{{\partial r^*}^2}-V_\ell^{\rm (even/odd)}(r) \right]
\Psi^{\rm (even/odd)}_{\ell m}(t,r) = S^{\rm (even/odd)}_{\ell m}(t,r) \,.
\eea
Here, $r^*=r+2M\ln[r/(2M)-1]$ with a BH's mass $M$,
and $V_\ell^{\rm (even/odd)}$ and $S^{\rm (even/odd)}_{\ell m}$
are the potential and source terms, respectively.
On the other hand, the NR waveforms are usually obtained from the NR $\psi_4$ data,
$\psi_4 = \ddot{h}_{+} - i\,\ddot{h}_{\times}$.
We should note that these are true only at $r=\infty$. 

In the analysis of the asymptotic behavior of the RWZ wave functions,
we have an asymptotic waveform~\cite{Lousto:2010qx,Nakano:2011pb},
\bea
\Psi_{\ell m}^{\rm (even/odd)}(t,r) &=& H_{\ell m} (t-r^*) 
+ \frac{\ell(\ell+1)}{2\,r} \int dt \, H_{\ell m} (t-r^*) 
+ O(r^{-2}) \,,
\label{eq:asymtRWZ}
\eea
for general ($\ell,\,m$) modes. Here, $H_{\ell m}$ is considered as the wave observed at infinity.
An error due to finite extraction radii
arises from the integral term on the right hand side of~(\ref{eq:asymtRWZ}).
Inverting the above relation, we have a perturbative extrapolation formula as
\bea
\fl
\left. \Psi_{\ell m}^{\rm (even/odd)}\right|_{r=\infty} &=& \Psi_{\ell m}^{\rm (even/odd)}(t,r)
- \frac{\ell(\ell+1)}{2\,r} \int dt \, \Psi_{\ell m}^{\rm (even/odd)}(t,r)
+ O(r^{-2}) \,. 
\eea
This expression is applied to waveforms
calculated in the black hole perturbation approach~\cite{Lousto:2010tb}.

Next, we discuss the mode function $\psi_4^{\ell m}$ of the Weyl scalar $\psi_4$.
If the NR Weyl scalar satisfies the Teukolsky equation~\cite{Teukolsky:1973ha}
in the Schwarzschild spacetime, 
this mode function is described by
\bea
r\,\psi_4^{\ell m}(t,r) &=& \ddot {\tilde H}_{\ell m} (t-r^*) 
+ \frac{(\ell -1)(\ell +2)}{2\,r} \dot {\tilde H}_{\ell m} (t-r^*)
+ O(r^{-2}) \,,
\label{eq:asymtPsi4}
\eea
where the difference between ${\tilde H}_{\ell m}$ 
and $H_{\ell m}$ in~(\ref{eq:asymtRWZ}) is just a numerical factor. 
Here, we have used the Kinnersley tetrad
to define the Weyl scalar $\psi_4$ from the Weyl tensor.
Inverting the above relation, we obtain a perturbative extrapolation formula
for $\psi_4^{\ell m}$ as
\bea
\left. r\,\psi_4^{\ell m}\right|_{r=\infty} &=& r\,\psi_4^{\ell m}(t,r)
- \frac{(\ell -1)(\ell +2)}{2\,r} \int dt \,[r\,\psi_4^{\ell m}(t,r)]
+ O(r^{-2}) \,.
\label{eq:original}
\eea
This is used for extrapolating gravitational waveforms in NR.

%%%%%%%%%%%%%%%%%%%%%%%%
\subsection{Phase and amplitude corrections}

To see the phase and amplitude corrections by use of the above formula,
we assume the decomposition of $H_{\ell m} $ in~(\ref{eq:asymtRWZ}) as
\bea
H_{\ell m} (t-r^*) = A_{\ell m} \exp(-i \omega_{\ell m} (t-r^*)) \,,
\label{eq:assumptionH}
\eea
with the amplitude ($A_{\ell m}$) and frequency ($\omega_{\ell m}$)
which are assumed to be real-valued and time-independent,
and the phase is given as $\phi_{\ell m}=\omega_{\ell m} (t-r^*)$.
Then, substituting (\ref{eq:assumptionH}) into~(\ref{eq:asymtRWZ}),
the RWZ wave functions at a finite extraction radius are written as
\bea
\fl
\Psi_{\ell m}^{\rm (even/odd)}(t,r) &=& A_{\ell m}
\left[1 + \frac{i \ell (\ell+1)}{2 \omega_{\ell m} r} \right] \exp(-i \omega_{\ell m} (t-r^*))
+ O(r^{-2})
% \cr \fl 
% &=& A_{\ell m}
% \sqrt{1 + \left(\frac{\ell (\ell+1)}{2 \omega_{\ell m} r}\right)^2 } \exp(-i \omega_{\ell m} (t-r^*)) \exp(i\,\delta \phi_{\ell m})
% + O(r^{-2})
\cr \fl 
&=& \left( A_{\ell m} + \delta A_{\ell m} \right) \exp(-i \omega_{\ell m} (t-r^*)) \exp(i\,\delta \phi_{\ell m}) + O(r^{-2}) \,,
\label{eq:APcollection}
\eea
where $\delta A_{\ell m}$ and $\delta \phi_{\ell m}$ are defined as
\bea
\frac{\delta A_{\ell m}}{A_{\ell m}} &=&
\frac{1}{2}\left(\frac{\ell (\ell+1)}{2 \omega_{\ell m} r}\right)^2 + O(r^{-4})\,,
\cr
\sin \delta \phi_{\ell m} &=& 
% \left. \left( \frac{\ell (\ell+1)}{2 \omega_{\ell m} r} \right) 
% \right/
% \sqrt{1 + \left(\frac{\ell (\ell+1)}{2 \omega_{\ell m} r}\right)^2 } = 
\frac{\ell (\ell+1)}{2 \omega_{\ell m} r} + O(r^{-2}) \,.
\eea
The phase correction from the extrapolation formula
has $O(r^{-1})$, and is the most significant correction.
On the other hand, the amplitude correction will be $O(r^{-2})$
which we have ignored here.
When we include the $O(r^{-2})$ correction for the waveforms,
the amplitude will be modified.
Therefore, the above analysis on the amplitude correction
is not accurate for the coefficient.
The above $r$-dependence is consistent with~\cite{Hannam:2007ik,Boyle:2009vi},
and has also been observed in the black hole perturbation
approach~\cite{Sundararajan:2007jg,Burko:2010au}.
Although we considered the corrections for the RWZ wave functions here,
this analysis is also applicable to the Weyl scalar.

%%%%%%%%%%%%%%%%%%%%%%%%%%%%%%%%%%%%%%%%
\section{More analysis and improved formula}\label{sec:Lazarus}
%%%%%%%%%%%%%%%%%%%%%%%%%%%%%%%%%%%%%%%%

In the previous section, 
we have used  the Kinnersley tetrad to define
the Weyl scalar $\psi_4$ which is in the Teukolsky formalism.
On the other hand, the tetrad used in NR is different from this.
Using~(2.15) in~\cite{Campanelli:2005ia},
we discuss the tetrad dependence.

Assuming the peeling theorem ($\psi_4 = [r\psi_4]/r,\,\psi_3 = [r^2\psi_3]/r^2,\,\psi_2 = [r^3\psi_2]/r^3,\,
\psi_1 = [r^4\psi_1]/r^4,\,\psi_0 = [r^5\psi_0]/r^5$,
where the functions in the square bracket are $O(r^0)$ for large $r$),
at a finite radius we have
\bea
\fl
r\psi_4^{\rm Kin} &=& 
\frac{1}{2}\,[r\psi_4^{\rm NR}]-{\frac {M [r\psi_4^{\rm NR}]}{{r}}}
-{\frac {a \left( 7\,a [r\psi_4^{\rm NR}] \, \cos^{2} \theta 
-3\,a [r\psi_4^{\rm NR}]  \right) }{4 {r}^{2}}}
\cr \fl &&
+i \left( {\frac {a\cos \theta \, [r\psi_4^{\rm NR}]}{{r}}}
-{\frac {a \left(  [r^2\psi_3^{\rm NR}] \, \sin \theta
+2\,M \,[r\psi_4^{\rm NR}] \,\cos \theta \right) }{{r}^{2}}} \right) 
+ O(r^{-3})\,,
\eea
where $\psi_4^{\rm Kin}$ and $\psi_4^{\rm NR}$ are
for the Kinnersley and NR tetrads, respectively.
The original correction (see, e.g.~\cite{Lousto:2010qx}) of the tetrad difference
in the extrapolation formula is just
\bea
r\psi_4^{\rm Kin} &=& 
\frac{1}{2}\,[r\psi_4^{\rm NR}]-{\frac {M [r\psi_4^{\rm NR}]}{{r}}}
= \left(\frac{1}{2}-\frac{M}{r}\right)\,[r\psi_4^{\rm NR}] \,.
\eea
Here, in order to improve the original formula,
we focus only on the $O(r^{-1})$ correction,
\bea
r\psi_4^{\rm Kin} &=& 
\frac{1}{2}\,[r\psi_4^{\rm NR}]-{\frac {M [r\psi_4^{\rm NR}]}{{r}}}
+ {\frac {i\,a [r\psi_4^{\rm NR}] \,\cos \theta }{{r}}} + O(r^{-2}) \,,
\label{eq:a-dep}
\eea
and consider the effect of the Kerr ($a$) term in the perturbative extrapolation.

It should be noted that we also take care of the $a$ dependence
in the Teukolsky formalism.
The wave function in the Teukolsky equation is
${}_{-2}\Psi=(r-i a \cos \theta)^4 \psi_4^{\rm Kin}$, while we simply used
${}_{-2}\Psi=r^4 \psi_4^{\rm Kin}$ by ignoring the $a$ dependence
in the original analysis from the RWZ equations~\cite{Lousto:2005xu}.
Including the $a$ dependence, we have
\bea
{}_{-2}\Psi &=& (r-i a \cos \theta)^4 
\left(
\frac{1}{2}\,\psi_4^{\rm NR}-{\frac {M \psi_4^{\rm NR}}{{r}}}
+ {\frac {i\,a  \psi_4^{\rm NR} \, \cos \theta}{{r}}} \right)
+ O(r^1)
\cr
&=&
\frac{r^4}{2}\,\psi_4^{\rm NR}-M \,r^3 \psi_4^{\rm NR}
- i\,a \, r^3  \psi_4^{\rm NR} \,\cos \theta + O(r^1)
\,,
\label{eq:-2psi_1}
\eea
from the difference in the tetrads, 
On the other hand, the radial Teukolsky equation
(with the Kinnersley tetrad) gives the asymptotic form of the mode function,
\bea
\fl 
{}_{-2}\Psi_{\ell m} &=&
\ddot {\tilde H}_{\ell m} (t-r^*) {r}^{3}
+ \biggl[ 
\frac{ \left( \ell-1 \right)  \left( \ell+2 \right)}{2} \dot {\tilde H}_{\ell m} (t-r^*)
-{\frac {4\,i\, ma}{\ell \left( \ell+1 \right) }} \ddot {\tilde H}_{\ell m} (t-r^*) \biggr] {r}^{2}
\cr \fl && + O(r^1,\,(a\omega)^2) \,.
\label{eq:-2psi_2}
\eea
In the above calculation, we have assumed the frequency domain waveforms
in~(\ref{eq:-2Pom}) with a monotonic frequency, for example,
$\dot {\tilde H}_{\ell m} (t-r^*)=-i \omega {\tilde H}_{\ell m} (t-r^*)$
in the transformation from the frequency to time domain analysis.
This assumption works well in the case of quasicircular orbits
because we consider the instantaneous frequency as the monotonic frequency.
The separation constant $\lambda$ is given in~(\ref{eq:approxL}) for $a\omega \ll 1$,
and we have ignored the $O((a\omega)^2)$ term.

Comparing (\ref{eq:-2psi_1}) with (\ref{eq:-2psi_2}) in each ($\ell,\,m$) mode,
\bea
\fl \ddot {\tilde H}_{\ell m} (t-r^*) 
+
\frac{\left( \ell-1 \right)  \left( \ell+2 \right) }{2\,r}\,\dot {\tilde H}_{\ell m} (t-r^*)
-{\frac {4\,i\, ma}{\ell \left( \ell+1 \right)\,r }} \ddot {\tilde H}_{\ell m} (t-r^*)
\cr
\fl \qquad
= 
\frac{1}{2}\left[ \left(1-{\frac {2M}{{r}}}\right) [r\psi_{4\ell m}^{\rm NR}(t,r)] 
- \frac{2\,i\,a}{r} 
\sum_{\ell' m'} C_{\ell m}^{\ell' m'}[r\psi_{4\ell' m'}^{\rm NR}(t,r)]\right] 
+ O(r^{-2},\,(a\omega)^2) \,,
\label{eq:-2psi_3}
\eea
where we have considered that the NR waveform is obtained
as $\psi_{4\ell m}^{\rm NR}(t,r)$ in
\bea
r\psi_4^{\rm NR} = \sum_{\ell m} [r\psi_{4\ell m}^{\rm NR}(t,r)] {}_{-2}Y_{\ell m}(\Omega)
\,.
\eea
The coefficients of mode couplings, $C_{\ell m}^{\ell' m'}$
which has a value for $\ell'=\ell$ and $\ell'=\ell \pm 1$ with $m'=m$,
are calculated as
\bea
C_{\ell m}^{\ell m} &=&
\int d\Omega {}_{-2}Y_{\ell m}^*(\Omega) \cos \theta {}_{-2}Y_{\ell m}(\Omega)
= \frac{2m}{\ell (\ell+1)} \,,
\cr
C_{\ell m}^{\ell+1\, m} &=&
\int d\Omega {}_{-2}Y_{\ell m}^*(\Omega) \cos \theta {}_{-2}Y_{\ell+1\, m}(\Omega)
\cr 
&=&
\frac{1}{\ell+1}
\sqrt {{\frac {  \left( \ell-1 \right) \left( \ell+3 \right)  \left( \ell-m+1 \right)  \left( \ell+m+1 \right)  }
{  \left( 2\,\ell+1 \right) \left( 2\,\ell+3 \right) }}} \,,
\label{eq:Ccoeff}
\eea
(see also Appendix A of~\cite{Berti:2014fga}).
It is noted that 
$r\psi_4^{\rm NR}$ at infinity is equal to $2 \ddot {\tilde H}_{\ell m} (t-r^*)$,
and if we consider only the $(\ell'=\ell,\,m'=m)$ coupling,
the $a$ term contributions cancel out in~(\ref{eq:-2psi_3}).
In the following analysis, we discard the factor $1/2$
in the right hand side of~(\ref{eq:-2psi_3})
due to the difference of the tetrads at infinity,
if necessary, this can be introduced again.
This means that we treat
\bea
\fl \ddot {\tilde H}_{\ell m} (t-r^*) 
+
\frac{\left( \ell-1 \right)  \left( \ell+2 \right) }{2\,r}\,\dot {\tilde H}_{\ell m} (t-r^*)
-{\frac {4\,i\, ma}{\ell \left( \ell+1 \right)\,r }} \ddot {\tilde H}_{\ell m} (t-r^*)
\cr
\fl \qquad
= 
\left(1-{\frac {2M}{{r}}}\right) [r\psi_{4\ell m}^{\rm NR}(t,r)] 
- \frac{2\,i\,a}{r} 
\sum_{\ell' m'} C_{\ell m}^{\ell' m'}[r\psi_{4\ell' m'}^{\rm NR}(t,r)]
+ O(r^{-2},\,(a\omega)^2) \,.
\label{eq:-2psi_3c}
\eea

We invert the above equation as
\bea
\fl \ddot {\tilde H}_{\ell m} (t-r^*) 
= 
\left(1-{\frac {2M}{{r}}}\right) [r\psi_{4\ell m}^{\rm NR}(t,r)] 
- \frac{2\,i\,a}{r} 
\sum_{\ell' m'} C_{\ell m}^{\ell' m'}[r\psi_{4\ell' m'}^{\rm NR}(t,r)]
\cr
\fl \qquad
-
\frac{\left( \ell-1 \right)  \left( \ell+2 \right) }{2\,r}\,\dot {\tilde H}_{\ell m} (t-r^*)
+{\frac {4\,i\, ma}{\ell \left( \ell+1 \right)\,r }} \ddot {\tilde H}_{\ell m} (t-r^*)
+ O(r^{-2},\,(a\omega)^2) 
\cr
\fl \qquad
= \left(1-{\frac {2M}{{r}}}\right) [r\psi_{4\ell m}^{\rm NR}(t,r)] 
- \frac{2\,i\,a}{r} 
\sum_{\ell' m'} C_{\ell m}^{\ell' m'}[r\psi_{4\ell' m'}^{\rm NR}(t,r)]
\cr
\fl \qquad
-
\frac{\left( \ell-1 \right)  \left( \ell+2 \right) }{2\,r}\,\int dt [r\psi_{4\ell m}^{\rm NR}(t,r)] 
+{\frac {4\,i\, ma}{\ell \left( \ell+1 \right)\,r }}  [r\psi_{4\ell m}^{\rm NR}(t,r)] 
+ O(r^{-2},\,(a\omega)^2) 
\cr
\fl \qquad
= \left(1-{\frac {2M}{{r}}}\right) [r\psi_{4\ell m}^{\rm NR}(t,r)] 
- \frac{2\,i\,a}{r} 
\sum_{\ell' \neq \ell} C_{\ell m}^{\ell' m}[r\psi_{4\ell' m}^{\rm NR}(t,r)]
\cr
\fl \qquad
-
\frac{\left( \ell-1 \right)  \left( \ell+2 \right) }{2\,r}\,\int dt [r\psi_{4\ell m}^{\rm NR}(t,r)] 
+ O(r^{-2},\,(a\omega)^2)
\,,
\eea
where we have considered the iteration procedure which is 
to substitute $\ddot {\tilde H}_{\ell m} (t-r^*) =[r\psi_{4\ell m}^{\rm NR}(t,r)]$
for the $O(r^{-1})$ terms, and used (\ref{eq:Ccoeff}) for the derivation of the third equality.
Finally, we have an improved extrapolation formula
\bea
\fl 
\left. r\,\psi_4^{\ell m}\right|_{r=\infty}
&=&
\left(1-{\frac {2M}{{r}}}\right) \left( r\psi_{4\ell m}^{\rm NR}(t,r) 
- \frac{(\ell -1)(\ell +2)}{2\,r} \int dt [r\psi_{4\ell m}^{\rm NR}(t,r)] 
\right)
\cr \fl &&
- \frac{2\,i\,a}{r} 
\sum_{\ell' \neq \ell} C_{\ell m}^{\ell' m}[r\psi_{4\ell' m}^{\rm NR}(t,r)] 
+ O(r^{-2},\,(a\omega)^2) \,.
\label{eq:improved}
\eea
Here, the factor $(1-2M/r)$ in the right hand side of the above equation
has been introduced in~\cite{Lousto:2010qx} to adjust the tetrad difference,
and is consistent with the analysis ignoring $O(r^{-2})$ in this section.

%%%%%%%%%%%%%%%%%%%%%%%%%%%%%%%%%%%%%%%%
\section{Discussion}\label{sec:diss}
%%%%%%%%%%%%%%%%%%%%%%%%%%%%%%%%%%%%%%%%

We have improved the perturbative extrapolation formula based on the Teukolsky formalism
to include spinning effects in the extraction background.
The improved formula in~(\ref{eq:improved})
is slightly more complicated than the original formula in~(\ref{eq:original}).
The original formula will give a good result for
the dominant $(\ell=2,\,m=\pm 2)$ mode of the waveform
because there are only sub-dominant modes
in the coupling due to the Kerr ($a$) term.

On the other hand, we should be careful when
we treat the sub-dominant modes.
For example, the $(\ell=3,\,m=\pm 2)$ mode has a contribution from
the dominant $(\ell=2,\,m=\pm 2)$ mode through the coupling.
And also, we need to choose properly the mass ($M$) and Kerr ($a$)
parameters in the improved formula of~(\ref{eq:improved}).
Since the mass and spin contributions are already in a higher order
with respect to $1/r$, we may use a simple estimation, 
$M=m_1+m_2$ and $a=(S_1 + S_2 + L_{\rm orb})/M$
for the aligned/anti-aligned cases
where $m_1$ and $m_2$ are the mass of each BH,
$S_1$ and $S_2$ are the spin,
and $L_{\rm orb}=(m_1m_2/M) \sqrt{M r_{\rm orb}}$
is the (Newtonian) orbital angular momentum, respectively.
At first glance, $L_{\rm orb}$ diverges for large orbital separation $r_{\rm orb}$,
but the correction of the $a/r$ term in (\ref{eq:improved}) is still small
because the extraction radius should be in the wave zone, {\it i.e.},
$r \gg \sqrt{r_{\rm orb}^3/M}$.
For precessing binary cases, we will need some complicated procedure
to include the Kerr term.
In the co-precessing frame (see, {\it e.g.}, 
\cite{Schmidt:2010it,O'Shaughnessy:2011fx,Boyle:2011gg}),
one of the candidate of $a$ is $a=|{\bf S}_1 + {\bf S}_2 + {\bf L}_{\rm orb}|/M$.
More practical procedure should be developed for precessing cases.
Also, it should be noted that we have assumed the monotonicity
of the frequency in the transformation from the frequency to
time domain analysis. The validity of this assumption should be
investigated.
Recently, (\ref{eq:improved})
has been studied by setting $a=0$ in~\cite{Nakano:2015pta}.

%%%%%%%%%%%%%%%%%%%%%%%%%%%%%%%%%%%%%%%%
\ack
%%%%%%%%%%%%%%%%%%%%%%%%%%%%%%%%%%%%%%%%

H.N. would like to thank C.~O.~Lousto and M.~Shibata, T.~Tanaka and K.~Kyutoku
for useful comments and suggestions.
H.N. acknowledges support by the Grant-in-Aid for 
Scientific Research No.~24103006,
and very fruitful and constructive referees' comments.

%%%%%%%%%%%%%%%%%%%%%%%%%%%%%%%%%%%%%%%%
\appendix

%%%%%%%%%%%%%%%%%%%%%%%%%%%%%%%%%%%%%%%%
\section{Estimation of the radiated energy, angular momentum and linear momentum}\label{sec:Lazarus_E}
%%%%%%%%%%%%%%%%%%%%%%%%%%%%%%%%%%%%%%%%

To find the correction in the radiated energy calculation,
we first extend the asymptotic waveform to the next order ($r^{-2}$)
in the Kerr background spacetime with mass $M$ and Kerr parameter $a$.
For simplicity, the frequency domain analysis is employed here.
The mode function of the Teukolsky function,
${}_{-2}\Psi=(r-i a \cos \theta)^4 \psi_4$, is obtained as
\bea
\fl
{}_{-2}\Psi_{\ell m \omega}(r) &=&
\biggl[( {r}^{3}+ i\, \left( m a+\frac{1}{2}\,{\frac {\lambda}{\omega}} \right) {r}^{2}
+ \biggl( \frac{1}{2}\,i \left( -3\,ia+i{m}^{2}a+2\,Mm \right) a
\cr 
\fl &&
+\frac{1}{2}\,{\frac {i \left( i\lambda\,m a + 3\,i m a + 3\,M \right) }{\omega}}
-\frac{1}{8}\,{\frac {\lambda\, \left( \lambda+2 \right) }{{\omega}^{2}}} \biggr) r 
+ O(r^0) \biggr] H_{\ell m \omega} \,,
\label{eq:-2Pom}
\eea
where $H_{\ell m \omega}/r$ becomes the waveform at infinity, 
and $\lambda$ is the separation constant in the radial and angular
Teukolsky equations~\cite{Teukolsky:1973ha}, for example,
we have the ($a\omega \ll 1$)-expression~\cite{Mano:1996vt,Sago:2005fn} as
\bea
\fl
\lambda &=& 
\left( \ell+2 \right)  \left( \ell-1 \right) 
-{\frac {2\,m \left( {\ell}^{2}+\ell+4 \right)}{\ell \left( \ell+1 \right) }} a\, \omega
+ \left( {\cal H} (\ell+1)- {\cal H} (\ell) \right) {a}^{2}{\omega}^{2}
+ O((a\omega)^3) \,;
\cr \fl &&
{\cal H} (\ell) = 2\,{\frac { \left( \ell-m \right)  \left( \ell+m \right) 
\left( \ell-2 \right) ^{2} \left( \ell+2 \right) ^{2}}{ \left( 2\,\ell-1 \right) {\ell}^{3} \left( 2\,\ell+1 \right) }} \,.
\label{eq:approxL}
\eea

The energy flux from the above asymptotic waveform is obtained as
\bea
\dot E_{\ell m \omega}(r) = \left(1 + \frac{6a\omega(a\omega-m)-\lambda}{2 \omega^2 r^2} + O(r^{-3}) \right)
\dot E_{\ell m \omega}^\infty \,,
\label{eq:Burko-Hughes}
\eea
via the square of the time integration of ${}_{-2}\Psi_{\ell m \omega}$.
Here $\dot E_{\ell m \omega}^\infty$ is evaluated from
the waveform at infinity, $H_{\ell m \omega}/r$. 
The above expression is the same as~\cite{Burko:2010au}
via the Sasaki-Nakamura equation~\cite{Sasaki:1981sx}.
In a similar way, we have a formula at a finite radius
for the radiated angular momentum~\cite{Campanelli:1998jv} as
\bea
\dot L_{\ell m \omega}(r) = \left(1 + \frac{6a\omega(a\omega-m)-\lambda}{2 \omega^2 r^2} + O(r^{-3}) \right)
\dot L_{\ell m \omega}^\infty \,.
\label{eq:Burko-Hughes-L}
\eea
The difference between $\dot E_{\ell m \omega}^\infty$ and $\dot L_{\ell m \omega}^\infty$
is the factor of $m/\omega$ in the frequency domain analysis.

When we discuss the time evolution of linear momentum of binaries,
the situation is slightly more complicated. Here, we focus only
on the radiated linear momentum along the $z$-axis.
In this case, we need to consider the mode coupling between different $\ell$ modes
in same $m$ modes. Formally (see, e.g.~\cite{Nakano:2010kv} for a detailed analysis
in the RWZ formalism)
\bea
\fl
\dot P_z(r) = \sum_{\ell \ell' m} 
\left(1+{\frac {i ( \lambda-\lambda' ) }{2 \omega\,r}}
+ \frac{ 24\,{a} \omega (a\omega - m) 
- (\lambda-\lambda' )^2 - 2 (\lambda - \lambda' ) }
{8 \omega^2 r^{2}} \right) \dot P_{z,\,\ell \ell' m}^\infty \,,
\eea
where the separation constant $\lambda'$ is associated with $\ell'$.
In contrast to the radiated energy and angular momentum,
there is an $O(r^{-1})$ correction in the calculation
of the radiated linear momentum.

%%%%%%%%%%%%%%%%%%%%%%%%%%%%%%%%%%%%%%%%
\section*{References}
%%%%%%%%%%%%%%%%%%%%%%%%%%%%%%%%%%%%%%%%

\bibliographystyle{iopart-num}
%\bibliography{../../../Bibtex/references}
\bibliography{./extrapolation.bbl}

%%%%%%%%%%%%%%%%%%%%%%%%%%%%%%%%%%%%%%%%
\end{document}